\documentclass[12pt]{article}
\usepackage{epsf}
\usepackage{amsmath}
\usepackage{graphics}
\usepackage{cite}

\renewcommand{\thefootnote}{\#\arabic{footnote}}
\begin{document}

\newcommand{\gtrsim}{ \mathop{}_{\textstyle \sim}^{\textstyle >} }
\newcommand{\lesssim}{ \mathop{}_{\textstyle \sim}^{\textstyle <} }

\newcommand{\rem}[1]{{\bf #1}}

\renewcommand{\thefootnote}{\fnsymbol{footnote}}
\setcounter{footnote}{0}
\begin{titlepage}

\def\thefootnote{\fnsymbol{footnote}}

\begin{center}
\hfill April 2013\\
\vskip .5in
\bigskip
\bigskip
{\Large \bf Cyclic Cosmology from the Little Rip}

\vskip .45in

{\bf Paul H. Frampton and Kevin J. Ludwick}

{\em Department of Physics and Astronomy,}

{\em University of North Carolina at Chapel Hill, NC 27599, USA}

\end{center}

\vskip .4in
\begin{abstract}
We revisit a cyclic cosmology scenario proposed
in 2007 to examine whether its hypotheses can be sustained if the
underlying big rip evolution, which was assumed there, is replaced
by the recently proposed little rip.
We show that the separation into causal patches at turnaround
is generally valid for a little rip, and therefore
conclude that the little rip is equally as suitable a basis for 
cyclicity as is the big rip.
\end{abstract}
\end{titlepage}

\renewcommand{\thepage}{\arabic{page}}
\setcounter{page}{1}
\renewcommand{\thefootnote}{\#\arabic{footnote}}

\newpage

A few years ago \cite{BF}, it was suggested, based on a suitably
modified big-rip type of
future evolution \cite{caldwell}, that a separation into
causal patches at turnaround and the subsequent contraction 
of an empty (except for dark energy) such patch
with zero entropy can lead to a cyclic cosmology.
In particular, the reduction to zero of the patch entropy
could avoid conflict with the second law
of thermodynamics. At the same time, and equally
important,
it could underly the low or vanishing entropy at
the beginning of inflation in the next expansion era.

One of the oldest questions in theoretical cosmology is whether
an infinitely oscillatory universe which avoids an initial singularity
can be consistently constructed. As realized by Friedmann\cite{Friedmann}
and especially by Tolman\cite{Tolman,TolmanBook} one principal
obstacle is the second law of thermodynamics which dictates that
the entropy increases from cycle to cycle. If the cycles 
thereby become longer, extrapolation into the past
will lead back to an initial singularity again, thus removing the
motivation to consider an oscillatory universe in the first place.
This led to the abandonment of the oscillatory universe by
the majority of workers.

In the present article, we examine whether this type
of cyclicity can be obtained in the little-rip
scenario of future cosmic evolution
\cite{FLS1} (see also \cite{FLNOS,FLS2}), focusing
at the beginning on the first model in \cite{FLS1}, although
our conclusion will apply to all little rip models.

An infinitely oscillatory universe is 
a very attractive alternative
to the big bang. One new ingredient in the cosmic make-up
is the dark energy discovered only in 1998, and so
it is natural to ask whether this can avoid the difficulties with entropy.

Some work has been done on the exploitation of the dark energy
in allowing cyclicity possibly without 
the need for inflation in
\cite{Ekpyrotic,SteinhardtTurok,SteinhardtTurok2,Boyle,SteinhardtTurok3}.
Another approach is the use of branes and a fourth spatial
dimension as in \cite{Sundrum,Randall,Binetruy,Freese}, which
examined consequences for cosmology. The big rip
and replacement of dark energy by modified gravity were
explored in \cite{PHFTT,PHFTT2}.

If the dark energy has a constant super-negative equation of state,
$\omega_{DE} = p_{DE}/\rho_{DE} < -1$, it leads
to a big rip\cite{caldwell} at a finite future time where there
exist extraordinary conditions with regard to
density and causality as one approaches the rip.  
In \cite{BF}, it was shown that these exceptional
conditions can assist in providing an infinitely cyclic model. 
We consider here a different model where $\omega_{DE} (a)$ evolves
with scale factor and where 
$\omega_{DE} \rightarrow -1$ 
asymptotically as $t \rightarrow \infty$,
leading to a a little rip. As we approach the little rip, 
expansion stops 
due to a brane contribution
and there is a turnaround
at time $t=t_T$ when the scale factor is deflated to 
a very tiny fraction ($f$) 
of itself and only one causal patch is retained, while
the other $1/f^3$ patches contract independently to
separate universes.
Turnaround takes
place at a time 
when the universe is 
fractionated into many independent
causal patches \cite{PHFTT2}. 

Contraction
occurs with a very much smaller universe than in
expansion and with vanishing entropy
because the universe is assumed empty of dust, matter
and black holes.

A bounce takes place 
a short time before a would-be big bang. After the bounce, entropy
is injected by inflation\cite{Guth}, where
it is assumed that an inflaton field is excited.
Inflation is thus a part of the present model which is
one distinction from the work of 
\cite{SteinhardtTurok,SteinhardtTurok2,Boyle,SteinhardtTurok3}.  
For cyclicity of the entropy, $S(t) = S(t + \tau)$, consistency with thermodynamics 
requires that the deflationary decrease by $f^3$ 
compensate the entropy increase acquired during
expansion, including the increase
during inflation.

%\bigskip

%\newpage

Let us review the Friedmann equation for the expansion phase.
Let the period of the universe be designated by $\tau$,
and let the bounce take place at $t = 0$ 
and turnaround at $t = t_T$.
Thus the expansion phase is for times $0 < t < t_T$ 
and the contraction phase corresponds to times
$t_T < t < \tau$.
We employ the following Friedmann equation for
the {\it expansion} period $0 < t < t_T$: 
\begin{equation}
\left( \frac{\dot{a}(t)}{a(t)} \right)^2  =   
\frac{8 \pi G}{3} \left[ \left( \rho_{DE} (t)
+\frac{\rho_{m_0}}{a(t)^{3}} +\frac{\rho_{r_0}}{a(t)^{4}}
\right)
-\frac{\rho_{total}(t)^2}{\rho_{crit}},
\right] 
\label{Friedmann}
\end{equation}
where the scale factor is normalized to $a(t_0)=1$ at the present time
$t = t_0 \simeq 14$ Gyr. 
$\rho_{i_0}$ denotes the value of the density component $\rho_i$
at time $t=t_0$. The first two terms are the contributions from dark energy and total matter
(dark plus luminous),
% satisfying
%\begin{equation}
%\Omega_{DE} = \frac{8 \pi G (\rho_{DE})_0}{3 H_0^2} = 0.72 
%~~~ {\rm and} ~~~
%\Omega_{m} = \frac{8 \pi G (\rho_m)_0}{3 H_0^2} = 0.28 
%\end{equation}
and $H_0 = \dot{a}(t_0)/a(t_0)$. The third term in the Friedmann equation
is the contribution from radiation.
% which is now 
%$\Omega_r = 1.3 \times 10^{-4}$. 
The final
term $\sim \rho_{total}(t)^2$ is derivable from a
brane setup\cite{Sundrum,Randall,Freese}; we use
a negative sign arising from negative brane tension (a negative sign
can arise also from a second timelike dimension, but that gives 
difficulties with closed timelike paths).  The constant critical density 
$\rho_{crit} = \rho_{total} (t_T) = \rho_{total} (\tau)$ is set appropriately 
so that the bounce occurs at the grand-unified-theory scale of our Universe, and
$\rho_{total} = \Sigma _{i=DE, m, r} ~\rho_{i}$.
As the turnaround is approached, the only significant terms 
in Eq. (\ref{Friedmann}) are the
first (where $\omega_{DE} < -1$) and the last. As the bounce is approached,
the only important terms in Eq. (\ref{Friedmann}) are the third and the last.
(In \cite{BF}, we argue that the second term, for matter, is absent during
contraction.)
In particular, the final term of Eq. (\ref{Friedmann}), $\sim \rho_{total}(t)^2$, 
arising from the brane setup is insignificant
for almost the entire cycle but becomes dominant
as one approaches $t \rightarrow t_T$ for the turnaround
and again for $t \rightarrow \tau$ approaching the bounce.

%\bigskip

At the turnaround, to sustain the scenario of \cite{BF}, we
must check whether the
causality structure is sufficiently close to that for the big rip.
We consider Model 1 of \cite{FLS1} and investigate this.  
The dark energy density we use for $a \geq 1$ is \cite{FLS1}
\begin{equation}
\label{littleripdensity}
\rho_{DE} (a) = \left(\frac{3 A \ln a}{2} + \rho_{DE_0}^{1/2} \right)^2   ,
\end{equation} 
where $A =3.46 \times 10^{-3} ~{\rm M}_{\odot}^{1/2}/{\rm pc}^{3/2}$.  
%and $\rho_{DE_0}$ is the present-day value of the dark energy density.  
For $a \leq 1$, $\rho_{DE} (a) = \rho_{DE_0}$ (the usual concordance model), which matches up 
with Eq. (\ref{littleripdensity}) at $a=1$.  

Quarks in a proton will begin 
to dissociate when $\rho_{DE} = \rho_{proton} = 7.8 \times 10^{69} ~{\rm M}_{\odot}/{\rm pc}^3$, or when 
$a = e^{2(\rho_{proton}^{1/2} - \rho_{DE_0}^{1/2} )/3A}$, assuming one fluid, the 
dominant dark energy.  We assume that the physical distance between 
two quarks in a proton before dissociation is the diameter of a proton, which is 1.8 fm.  
This evolution of distance between two such quarks is governed by the  
timelike geodesic equation with a source equal to the acceleration due to the strong force 
trying to hold together the two quarks:
\begin{equation}
\label{timelike}
\begin{split}
&\frac{1}{a} \frac{d\rho}{dt} \frac{d}{d\rho} \left(\frac{d\rho}{dt} \frac{dR}{d\rho} \right) \\
&+ \frac{R}{a} \frac{4 \pi G}{3} \left[ \rho - \frac{4 \rho^2}{\rho_{crit}} - 2 p \left( \frac{2 \rho}{\rho_{crit}} - 1 \right) \right] \\
&= \frac{2 \hbar c \alpha_s}{M_{q}} \left( \frac{1}{R^2} + \frac{1}{(6.48 \times 10^{-33} ~{\rm pc})^2} \right).
\end{split}
\end{equation}
The chosen critical density is $\rho_{crit} = 6.84 \times 10^{108} ~{\rm M}_{\odot}/{\rm pc}^3$, 
$R = a(t) r$ is the physical distance, the righthand side 
of Eq. (\ref{timelike}) is the experimentally determined expression for 
the proton's binding force\cite{strongforce} 
and $M_{q}$ is the mass of a down quark.  
  
Integrating the null geodesic equation, we see that light travels a comoving distance of 
\begin{equation}
\label{null}
\begin{split}
&\frac{c e^{2 \rho_{DE_0}^{1/2}/3A}}{2A \sqrt{6 \pi G}}  
\int_{\tilde{\rho}}^{\rho} \frac{ e^{- 2 \rho'^{1/2}/ 3A }}{\rho'^{1/2} (\rho' - \frac{\rho'^2}{\rho_{crit}} )^{1/2}} d\rho' \\
&= 3.22 \times 10^8 {\rm pc} ~[ \operatorname{Ei} (- 192.68 ~{\rm pc}^{3/2}/{\rm M}_{\odot}^{1/2} ~ \rho^{1/2}) - 
\operatorname{Ei} (- 192.68 ~{\rm pc}^{3/2}/{\rm M}_{\odot}^{1/2} ~ \tilde{\rho}^{1/2})] ,
\end{split}
\end{equation}
where $\operatorname{Ei} (x) = \int_{- \infty}^x \frac{e^t}{t} dt $ is the exponential integral function.

%By comparing the distances travelled by light and the dissociated quark from the time of proton dissociation until the turnaround, one 
%sees that the quark travels a greater physical distance, and so the 
%quarks of a proton are causally disconnected by the turnaround.  
By using Eqs. (\ref{timelike}) and (\ref{null}), one sees that at some time before the turnaround, 
the quarks from all protons are dissociated to the extent that they are causally disconnected.  
Structures that are less strongly bound will dissociate earlier 
since the inertial force due to dark energy increases monotonically, so these structures will 
also be causally disconnected by the turnaround.  Therefore, all bound structures will dissociate by the turnaround, and they will reside 
isolated in causally disconnected patches of spacetime.  
Other models in \cite{FLS1} all grow more quickly and 
dissociate bound structures sooner than 
Model 1 of \cite{FLS1} discussed here, so these models will also 
create causally disconnected patches of spacetime.  

%\bigskip
%\bigskip

%\noindent
From this result, it follows that a cyclic cosmology along the lines of
\cite{BF} can be sustained by a little rip equally as well as with a big rip.
In other words, the little rip can be used 
%, as can the pseudo rip \cite{FLS2},
as the basis for the turnaround between expansion
and contraction eras because
the causal patch structure is sufficiently similar to that for the big-rip case.

\bigskip

%\acknowledgements 
\bigskip

\begin{center}

{\bf Acknowledgements}

\end{center}

\bigskip
One of us (PHF) thanks G. Gelmini for a useful discussion.
This work was supported in part by the
U.S. Department of Energy under Grant No. DE-FG02-06ER41418.


\newpage

\bigskip

\begin{thebibliography}{100}
\bibitem{BF}
L. Baum and P.H. Frampton, Phys. Rev. Lett. {\bf 98,} 071301 (2007).
{\tt astro-ph/0610213}.
\bibitem{caldwell}
R.R. Caldwell, Phys. Lett. {\bf B545,} 23 (2002). {\tt astro-ph/9908168};
R.R. Caldwell, M. Kamionkowski and N.N. Weinberg,
Phys. Rev. Lett. {\bf 91,} 071301 (2003).
{\tt astro-ph/0302506}.
\bibitem{Friedmann}
A. Friedmann, Z. Phys. {\bf 10,} 377 (1922).
\bibitem{Tolman}
R.C. Tolman, Phys. Rev. {\bf 38,} 1758 (1931).
\bibitem{TolmanBook}
R.C. Tolman, Relativity, Thermodynamics and Cosmology. Oxford University Press (1934).
\bibitem{FLS1}
P.H. Frampton, K.J. Ludwick and R.J. Scherrer,
Phys. Rev. {\bf D84,} 063003 (2011).
{\tt arXiv:1106.4996[astro-ph.CO]}.
\bibitem{FLNOS}
P.H. Frampton, K.J. Ludwick, S. Nojiri, S.D. Odintsov and R.J. Scherrer,
Phys. Lett. {\bf B708,} 204 (2012).
{\tt arXiv:1108.0067[hep-th]}.
\bibitem{FLS2}
P.H. Frampton, K.J. Ludwick and R.J. Scherrer,
Phys. Rev. {\bf D85,} 083001 (2012).
{\tt arXiv:1111.2964[astro-ph.CO]}.
\bibitem{Ekpyrotic}
J. Khoury, B.A. Ovrut, P.J. Steinhardt and N. Turok.
Phys. Rev. {\bf D64,} 123522 (2001).
{\tt hep-th/0103238}.
\bibitem{SteinhardtTurok}
P.J. Steinhardt and N. Turok, Science {\bf 296,} 1436 (2002). {\tt hep-th/0111030}.
\bibitem{SteinhardtTurok2}
P.J. Steinhardt and N. Turok, Phys. Rev. {\bf D65,} 126003 (2002).
{\tt hep-th/0111098}.
\bibitem{Boyle}
L.A. Boyle, P.J. Steinhardt and N. Turok, Phys Rev. {\bf D70,} 023504 (2004)
{\tt hep-th/0403026}.
\bibitem{SteinhardtTurok3}
P.J. Steinhardt and N. Turok, Science {\bf 312,} 1180 (2006). 
{\tt astro-ph/0605173}.
\bibitem{Sundrum}
L. Randall and R. Sundrum, Phys. Rev. Lett. {\bf 83,} 3370 (1999). {\tt hep-ph/9905221}; {\it ibid}.
{\bf 83,} 4690 (1999). {\tt hep-th/9906064}.
\bibitem{Randall}
C. Csaki, M. Graesser, L. Randall and J. Terning, Phys. Rev. {\bf D62,} 045015 (2000).
{\tt hep-ph/9911406}.
\bibitem{Binetruy}
P. Bin\'{e}truy, C. Deffayet and D. Langlois, Nucl. Phys. {\bf B565,} 269 (2000).
{\tt hep-th/9905012}.
\bibitem{Freese}
M.G. Brown, K. Freese and W.H. Kinney, JCAP {\bf 0803}:  002, (2008). {\tt astro-ph/0405353}.
\bibitem{PHFTT}
P.H. Frampton and T. Takahashi, Phys. Lett. {\bf B557,} 135 (2003).
{\tt astro-ph/0211544}.
\bibitem{PHFTT2}
P.H. Frampton and T. Takahashi, Astropart. Phys. {\bf 22,} 307 (2004).
{\tt astro-ph/0405333}.
\bibitem{Guth}
A.H. Guth, Phys. Rev. {\bf D23,} 347 (1981).
\bibitem{strongforce}
E. Eichten {\it et al.}, Phys. Rev. Lett. {\bf 34,} 369 (1975). 

\end{thebibliography}
\end{document}